# Gamified Informed Decision-Making for Performance-Aware Design by Non-Experts: An Exoskeleton Design Case Study

Arman KHALILBEIGI KHAMENEH*[a], Armin MOSTAFAVI[b] , Alicia NAHMAD VAZQUEZ[a]

* University of Calgary, SAPL, Calgary, CANADA
arman.khalilbeigikha@ucalgary.ca

[a] University of Calgary, SAPL, Calgary, CANADA
[b] Human Centered Design Department, Cornell University, Ithaca, New York, USA

## Abstract

Decision Support Systems (DSS) play a crucial role in enabling non-expert designers to explore complex, performance-driven design spaces. This paper presents a gamified decision-making framework that integrates game engines with real-time performance feedback. Performance criteria include structural behavior, environmental parameters, fabrication, material, and cost considerations. The developed design framework was tested with architecture students and non-expert designers on the design of an exoskeleton facade to retrofit an existing building. Participants (N=24) were able to iteratively modify façade geometries while receiving real-time feedback across the three key criteria: 1) structural behavior, including deflection, mass, and stress/strength ratio; 2) environmental parameters, such as solar gain and heating/cooling energy demands; and 3) fabrication considerations, including fabrication and material costs, robotic machining, and material setup. The evaluation of participant interactions reveals that gamified feedback mechanisms significantly enhance user comprehension and informed decision-making across the criteria. Further, participants' understanding of structural, material, and fabrication performance in relation to the iterative design task suggests that curated design spaces and structured guidance improve efficiency compared to open-ended generative tools. This research contributes to pre-occupancy evaluations, demonstrating how gamified environments enable stakeholder participation in the design process through informed decision-making and customized negotiation of performance criteria. The study highlights the potential of game engines as mediators between computational analysis and non-expert designers or students in the early stages of their education, aiming to enhance their understanding of geometry, structural, environmental, and material performance. The framework offers scalable applications of performance-aware design in the early stages of architectural education.

## 1. Introduction

Fostering human potential in creativity, innovation, and performance remains a persistent challenge, especially in design fields, where supporting complex activities is essential for solving multifaceted problems [1]. In architectural design, the need for performance-driven approaches is growing, with designers required to address interrelated criteria such as structural integrity, environmental sustainability, and fabrication feasibility [2]. However, traditional pipelines often demand advanced expertise, creating barriers for non-expert designers and novices in related fields. To address this gap, this paper introduces a novel gamified decision-making framework that incorporates real-time performance feedback within a game engine environment, specifically tailored to empower non-expert users during the early stages of architectural design. By embedding gamification and serious game principles, the proposed framework transforms performance-driven design exploration into an intuitive and engaging process, aligned with human-centered design [3]. While decision support systems for design performance analysis have advanced, they remain largely expert-oriented, with interfaces and workflows inaccessible to novices. Early efforts at innovative interfaces, such as Architecture Machine and Flatwriter, anticipated the benefits of gamified strategies, but integrated and accessible solutions remain rare [4]. Recent research highlights the potential of gamification in enhancing motivation and learning outcomes in design education [5]. However, the field lacks cohesive frameworks that address simultaneously structural, environmental, and fabrication considerations for non-expert users. Persistent gaps include the absence of platforms synthesizing multiple performance criteria for novices, limited use of real-time feedback in iterative design, and a lack of gamified approaches balancing creative freedom with structured guidance.

## 1.1 Serious Games, Gamification, and Informed Decision-Making

Integrating game mechanics into design methodologies marks a significant evolution in contemporary design, allowing broader participation and enhancing the quality of decision-making, particularly for non-expert users [6]. While the adoption of game-based strategies in design is on the rise, the literature still lacks clarity regarding fundamental definitions and how concepts such as serious games, gamification, and informed decision-making are applied in practice [7]. In this context, serious games are understood as interactive, rule-based computer experiences with goals that extend beyond entertainment, providing continual feedback to promote learning and progress [8]. Gamification involves introducing game-like mechanics and thinking into non-game settings to improve engagement, motivation, and problem-solving [9]. Of particular relevance to this work is the notion of gamified informed decision-making, in which users make autonomous choices based on clear, accessible information, supported by an understanding of alternatives, potential outcomes, and personal or contextual goals [10]. Within the proposed framework, elements such as points, levels, challenges, storytelling, and instant feedback are not simply added, but are thoughtfully adapted to reinforce learning, structure complexity, and promote active exploration. Features including mini-tasks, visual cues, adaptive pathways, and digital guidance enable users to experiment, reflect, and iterate without overwhelming cognitive demands, fostering an environment where they are engaged, empowered, and able to make and iterate upon informed design decisions.

## 1.2 Non-expert Involvement in IDM and Design Tool

Recent advancements in Decision Support Systems (DSS) have brought data-driven insights to the forefront of architectural design, offering the potential to improve complex decision-making processes [11]. Despite these advances, existing methodologies often fall short when it comes to practical implementation and scalability in real-world building projects [12]. Systematic reviews underscore the growing importance of combining gamified and participatory approaches to broaden engagement beyond expert users [13]. This challenge is especially acute in the early design stages, where foundational decisions have the greatest downstream impact. Research shows that individuals often prefer qualitative reasoning over quantitative analysis when interacting with design information, particularly in initial phases [14], early decisions can account for up to 70% of a product's total cost [15]. Despite the potential for collaborative, user-centered interventions at this stage, most design tools remain fragmented and ineffective in engaging non-experts, hence also limiting stakeholder participation. These tools generally fall into three categories: CAD software, generative/algorithmic design tools, and analytical/simulation platforms. Each has its own limitations. CAD tools lack performance feedback, generative tools have steep learning curves with complex feedback mechanisms, if at all, and analytical tools disrupt workflows with external data dependencies, posing particular challenges for non-experts. High cognitive load, insufficient real-time feedback, opaque algorithms, and a misalignment between user mental models and tool design further limit the effectiveness of these platforms [16], [17]. Furthermore, fragmented workflows and a lack of motivational features hinder experimentation and reflective learning, making it difficult for novice users to build design expertise or experience a sense of accomplishment. These challenges underscore the need for integrated, user-centered frameworks that prioritize accessibility, real-time feedback, and learning support, empowering non-expert participation in complex design processes.

### 1.3 Structure, Energy, and Cost Associated with Façade Design

Façade design critically influences a building's structural integrity, energy performance, and overall cost, functioning as both an environmental barrier and a key visual interface [12]. In the context of retrofitting buildings, façades are crucial for enhancing both performance and aesthetics without compromising the more carbon-intensive core components, such as columns and slabs. With many downtown cores facing vacancies, retrofitting buildings offers a practical and low-cost strategy for revitalizing urban spaces. However, conventional design tools typically present performance data in highly technical terms, limiting accessibility for non-expert users, particularly in early design stages when decisions have the most significant long-term impact [18]. This creates a gap in holistic, user-centered approaches that can engage a diverse range of stakeholders in façade-related decision-making. Recent research highlights the need for multidisciplinary frameworks that integrate building science, environmental psychology, and human-computer interaction to better address user-facade interactions and support informed choices [12]. Responding to these challenges, this research proposes a gamified informed decision-making (IDM) framework that delivers real-time, accessible feedback on structural, environmental, and cost metrics for non-expert designers during façade design. By leveraging interactive, evidence-based feedback within a gamified environment, users can explore and balance trade-offs among structure, energy use, and cost without needing advanced technical expertise [14]. The findings indicate that immediate performance feedback can significantly enhance decision quality, with participants achieving up to 51% improvement across key metrics such as structural displacement and operational carbon. Furthermore, the system promotes micro-learning and consistent engagement, demonstrating that non-experts, when provided with actionable information, can contribute meaningfully to early-stage design and help produce more resilient, sustainable outcomes [19]. This paper further explores how real-time feedback on structure, energy, and cost influences user decision-making in façade design.

## 2. Methods

This study addresses these challenges by developing and evaluating a gamified framework that provides real-time feedback across key performance domains, enabling non-experts to navigate complex design spaces. By using game engines as mediators for performance-aware design, the research contributes to increasing stakeholder participation, design pedagogy, gamification, and human-centered computing. It demonstrates that gamified feedback enhances comprehension of complex criteria and offers a scalable foundation for incorporating performance-aware design skills into early architectural education and for non-expert scenarios. Ultimately, the study informs the development of future design tools that strike a balance between structured guidance and creative exploration.

### 2.1 Tool and Interface Development

The finalized façade was iteratively developed using a structured gamified framework designed to enable informed decision-making through real-time, performance-aware feedback in an immersive environment. The system architecture integrates generative design tools, performance simulation engines, and a high-fidelity interaction layer within Unreal Engine 5 (UE5). At its core, the framework consists of three modular layers: a backend implemented in Grasshopper™ for parametric geometry and Python for custom analytics, a communication mediator utilizing a custom bi-directional socket protocol, and a frontend interface constructed in Unreal Engine using Blueprint scripting and C++. The backend manages key computational tasks, including structural analysis, daylight and environmental simulations, and fabrication assessments, generating analytical data through tools compatible with Rhinoceros and custom modules that function as independent services. Outputs are organized and formatted by a data shield and Data I/O Manager, allowing the system to prioritize rapid generative feedback while deferring more computationally intensive tasks.

Optimization strategies, such as data dams within Grasshopper logic, ensure partial simulation results contribute to real-time feedback loops. The frontend hosts modules for experiment flow, participant input, backend synchronization, and behavioral tracking, translating analytics into intuitive, actionable metrics for non-expert users. This modular, distributed architecture enables iterative experimental changes without requiring full recompilation, thereby combining the strengths of each environment while ensuring scalability and extensibility. Maintaining uninterrupted, real-time engagement is a key design goal, as efficient data flow is crucial for supporting active user participation and rigorous experimental investigation during the early-stage façade design process.

## 2.2 Decision-Making Experiment Conditions

First, we set a range of experimental conditions to test in our experiment (Figure 1). The experiment employed a within-subjects design in which each participant engaged with two primary conditions:

*Informed Decision-Making (IDM) Condition:* In this condition, participants received real-time, encapsulated feedback on multiple performance metrics, specifically structure, environment (energy), and cost, while making design configuration choices (such as depth, width, number of layer laminations, and dimension of the exoskeleton facade modules). Feedback was provided immediately after each decision, allowing participants to iteratively explore the design space and understand the implications of their choices across all metrics.

*Non-Informed Decision-Making (nIDM) Condition:* In this control or baseline condition, participants made design decisions without receiving performance feedback. Here, users were still able to explore and modify design options, but they lacked access to real-time information about the structural, energy, or cost implications of their choices.

Participants completed tasks under both conditions in randomized order, allowing for direct comparison of their behavior, decision-making time, and the quality of their design outcomes with and without access to real-time performance feedback.

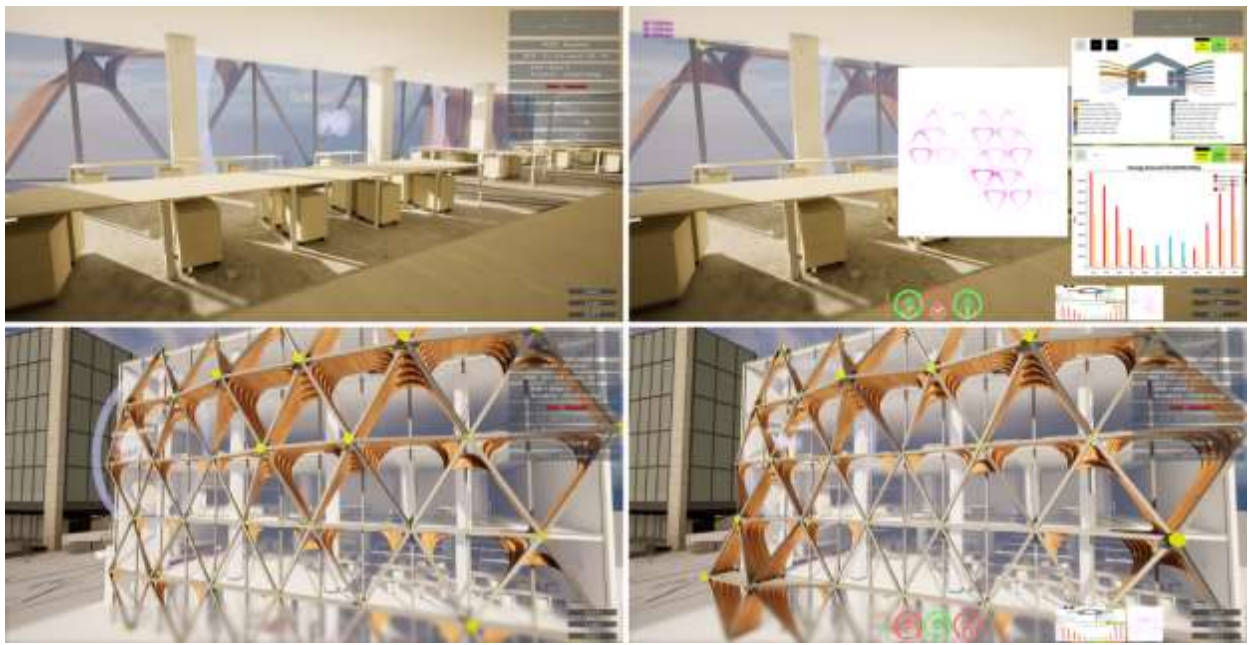

Figure 1: Decision-making conditions: Non-Informed Decision-Making condition (left),
Informed Decision-Making condition (right).

## 2.3 Participants

Twenty-four healthy adult participants were recruited through convenience sampling, using word-of-mouth and announcements on departmental email lists. The majority were affiliated with the University of Calgary. To ensure validity and repeatability, the study defined "early designer" with specific inclusion and exclusion criteria. Participants qualified if they had completed two years or more of formal education or training in design-related fields such as architecture. Individuals were excluded if they had six months or more of professional experience in roles involving building performance analysis, such as energy modeling or structural engineering. As a result, the participant pool consisted of non-expert designers, interdisciplinary collaborators, and emerging professionals from complementary fields, including building modelers, architects, landscape architects, and planning architects. Adherence to these criteria was ensured through a pre-screening process based on self-reported information and was further monitored by the author during data collection. Ethical approval for this study was obtained from the Conjoint Faculties Research Ethics Board (CFREB) at the University of Calgary with Ethics ID REB21-1935 REN1-2-3. All participants provided informed consent prior to participation in accordance with the approved protocol.

## 2.4. Measurements and Metrics

To comprehensively evaluate participant interactions and system usability, several measurement tools were employed throughout the study. The backend framework evaluated each design configuration across seven quantitative performance metrics, categorized under structure, environment, and fabrication. Structural performance included C1: maximum displacement (cm) at key points such as midpoints and endpoints under different load cases; C2: internal elastic energy (kNm), calculated as the scalar product of nodal loads and their displacements; and C3: total structural mass (kg), indicating the material requirements for the exoskeleton. These three metrics were computed using Finite Element Analysis (FEA) tools developed by Bollinger + Grohmann™ [20]. Environmental performance was assessed using early-stage simulation tools developed at ETH Zurich [21] and consisted of C4: annual operational cost (CAD/year) covering heating, cooling, electricity, and domestic hot water; C5: carbon emissions (kg $CO_2$/year), which included both annual operational emissions and annualized embodied emissions from construction; and C6: annual solar gain (kWh/year), reflecting the passive energy benefit received by the building envelope. The final metric, C7, represented fabrication complexity, a composite performance index developed by the authors to capture the constructability of a design. It combined material consumption (M, kg) and machining time (T, hours) through the formula: FC = $\omega_m$ × (M / M_ref) + $\omega_t$ × (T / T_ref), where $\omega_m$ and $\omega_t$ are weights (0.5 each in the main version), and M_ref and T_ref are the maximum reference values observed across all design options.

To ensure usability and maintain experimental integrity, invalid or non-constructible design options were automatically adjusted to the nearest valid alternative, preventing workflow disruptions and user confusion. While the alpha version displayed seven performance metrics as separate bar charts, users reported difficulty interpreting conflicting improvement options (e.g., lower mass vs. higher solar gain). In response, the beta version introduced a comparative feedback model, grouping metrics into three domains, structure, environment, and fabrication, and labeling each as improved, neutral, or worsened based on relative changes (named ENC1-3, respectively). An improvement of each label was defined as at least two of three sub-metrics trending positively within each domain. This tiered feedback system improved clarity, engagement, and supported both intuitive and expert-level decision-making.

All user actions and system interactions were logged with timestamps, enabling detailed analysis of decision-making patterns, pacing, and responses to feedback. The platform tracked the sequence and frequency of design modifications, capturing user exploration and adaptive behavior. Following task completion, participants completed the System Usability Scale (SUS) [22], providing standardized usability scores for benchmarking. These tools enabled a comprehensive evaluation of user experience, system effectiveness, and the influence of feedback on design behavior.

## 2.5 Procedure

Each session began with an interactive tutorial to familiarize participants with the interface and core mechanics. Through a series of short exercises, users learned to navigate the 3D space, select elements, and manipulate design configurations using intuitive controls. A brief summary video and a consolidation task followed the tutorial to reinforce understanding. Participants then completed two façade design tasks in a randomized within-subject order: one with real-time feedback (IDM condition) and one without (nIDM condition). To reduce confounding variables, interaction was limited to pushing or pulling nodes on the exoskeleton; all other changes were system-generated. Design adjustments were made using a 3D clicking interface with scrolling or keyboard input, and constructability was verified automatically. Invalid configurations were replaced with the nearest valid alternative.

In the IDM condition, participants received encapsulated feedback across three domains: structure, environment, and fabrication. Seven metrics were aggregated and visually encoded using color-coded icons indicating improvement, neutrality, or decline relative to the previous state. Additional visuals, such as mesh overlays and energy diagrams, were accessible on demand. In the nIDM condition, no feedback was provided; only geometric outcomes were visible. Sessions were conducted in a controlled lab setting, one participant at a time, with only the researcher present. Participants reviewed and signed informed consent before beginning. Each task was self-paced and completed upon user satisfaction or design selection. After both tasks, participants completed the SUS questionnaire and were invited to an optional exit interview. The system maintained high-level, intuitive interaction, while computational processes operated in the background to support non-expert usability.

## 3. Results

This section reports the findings based on user behavior recorded in the framework log. First, it presents time-related findings, establishing the effect of receiving feedback on decision-making time. This is followed by how receiving feedback has led to improvements in performance criteria and patterns of user engagement with the design subject.

### 3.1 Platform Usability and Decision-making Time

The post-session SUS results support the conclusion that participants engaged with the system in a deliberate and meaningful manner. The average SUS score was 81.3 (SD = 10.2), with a range from 52.5 to 97.5. These values significantly exceed the established benchmark of 68, situating the system within the top decile of evaluated interfaces and corresponding to an excellent rating on the adjective scale. The clustering of scores between 76.25 and 88.75 further reinforces the system's consistently high usability. Collectively, these outcomes suggest that the platform offers a robust and accessible user experience, particularly for individuals without prior design experience.

The effect of real-time feedback on decision-making time was evaluated across 4,292 design choices. Participants took an average of 5.33 seconds with feedback and 6.01 seconds without. Although the no-feedback condition showed a higher mean, its lower median (0.66 vs. 0.84 seconds) indicates the presence of outliers. Non-parametric tests confirmed a statistically significant difference between conditions (Mann-Whitney $U = 2,320,806.5$, $p = 0.0043$; Wilcoxon $= 570,990.5$, $p = 0.0010$), suggesting that real-time feedback supported more efficient (faster and more consistent) decision-making.

To examine learning effects, the slope of decision time across successive design attempts was calculated for each participant. Results showed a mean slope of –0.0235 (SD = 0.0878) for the feedback condition (IDM) and 0.0557 (SD = 0.1314) for the no-feedback condition (nIDM), with a mean difference of –0.0791 ($t = -1.8460$, $p = 0.0778$). While this did not reach conventional statistical significance, the trend suggests that feedback promoted faster decision-making over time. Notably, individual learning rates in the feedback condition ranged from moderate (–0.0379) to substantial (–0.4960), reflecting considerable variability in adaptive behavior (Figure 2).

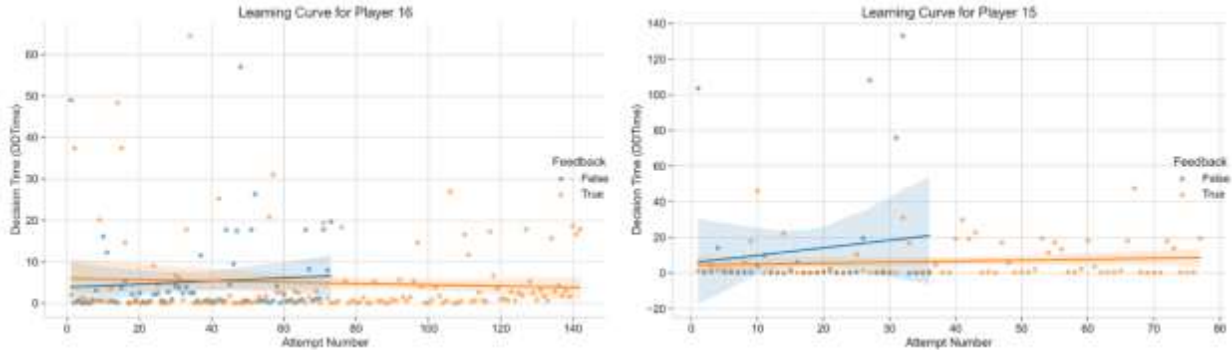

Figure 2: Micro-learning effect (mean slope comparison) of two sample participants in both conditions

### 3.2 Façade Performance Metrics in IDM vs nIDM

To assess the quality of design outcomes, the final three decisions made by each participant in both conditions were analyzed in relation to key performance metrics (Figure 3). The results demonstrate that real-time feedback consistently guided participants toward higher-performing choices across three encapsulated categories: structural efficiency (C1-C3), environmental performance (C4-C6), and fabrication complexity (C7). The data has 5% of outliers. Removing them, the data shows a clear downward trend in all three categories, indicating improved outcomes. Specifically, users in the feedback condition (IDM) made decisions that lowered structural and environmental loads while reducing fabrication complexity. On an individual level (comparing decisions of each user between conditions) examination participants showed an average mean change of -0.90%, -19.84%, -1.60%, -0.12%, -0.04%, 1.24%, and -2.65% in C1 to C7, respectively. This suggests that, on average, feedback had a positive effect on performance. At the group level (comparing the overall median behavior of all participants in each condition), sorting the seven metrics by the change in their mean value reveals that the top three metrics with the most changes among users are: 1. Cost (62.5% improvement), 2. Weight of the structure (58.3%), and 3. Solar gain (54.2%). The least changed is carbon emission (41.7%). Most users preferred the alignment of the structure and cost, indicating that they favored improvements in these two categories over the environment. 41.7% of participants improved in 5-7 measures. 20.8% of participants improved in 3-4 measures, and 37.5% of participants improved in 0-2 measures. These patterns confirm that participants not only engaged meaningfully with the system but also internalized the encapsulated feedback, applying it to optimize their design outcomes.

This outcome highlights a significant contribution of the framework, even without direct exposure to underlying metrics, participants used abstracted, labeled feedback to iteratively improve design quality. Together with earlier findings on decision time and learning effects, three core insights emerge: (1) real-time feedback reduced decision-making time, (2) micro-learning effects became increasingly evident over the course of interaction, and (3) participants responded meaningfully to feedback, achieving measurable improvements in unseen performance metrics through limited but well-structured guidance. This underscores the framework's potential to support non-expert users in performance-driven design tasks.

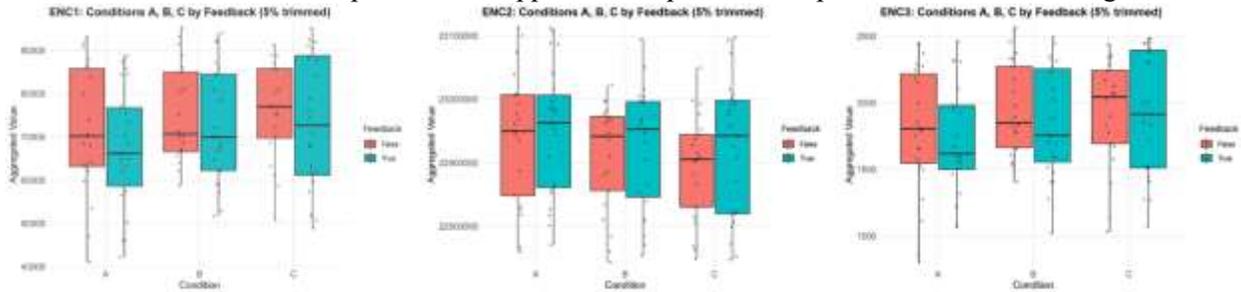

Figure 3: Performance metrics of the last three design choices (A, B, C = Final Configuration), ENC1 (C1 to C3 combined), ENC2 (C4 to C6 Combined), and ENC3 (C7)

A comparative analysis of the percentage change in performance metrics across all participants reveals the influence of real-time feedback on design behavior (Figure 4). Each row represents an individual participant, with color intensity reflecting the magnitude of deviation from baseline performance (defined as the overall average across all participants and conditions). The binary overlay further classifies outcomes: green cells indicate improved performance in the feedback condition, while white cells denote no improvement. A blue border marks participant whose final design outcome outperformed their baseline in the majority of measured categories. This visualization reveals a consistent pattern of enhanced performance following feedback, suggesting that participants not only engaged more effectively with the design interface but also internalized the encapsulated performance information, making more informed decisions. The fractional values on the right summarize the number of metrics (out of seven) in which each participant exceeded the baseline under the feedback condition.

While this visual representation supports the hypothesis that gamified, relative, and encapsulated feedback enhances performance outcomes, it is essential to recognize the methodological refinement applied in the analysis. Previous comparisons relied on aggregate means or percentage deviations. Whereas this study focuses exclusively on the final set of design decisions made by each participant in each condition. This approach accounts for the latent nature of learning and behavioral adaptation over time, providing a more precise assessment of feedback effectiveness. It demonstrates that even minimal, abstracted feedback can lead to measurable improvements in performance-driven design tasks.

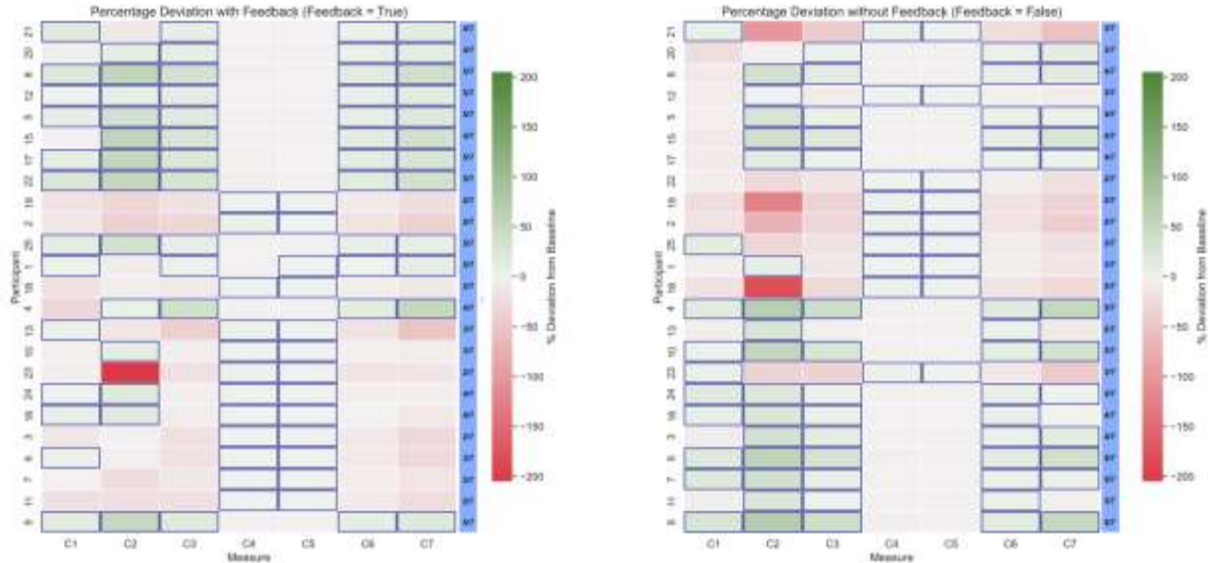

Figure 4: Percentage and binary deviation from baseline

### 3.3 Interaction and Exploration Behaviors

The influence of real-time feedback on users' spatial orientation and interaction patterns is illustrated in Figure 5, which maps focal points for each participant across conditions. The spherical representation reflects each user's field of view. When participants stood inside the building and looked outward, back toward the façade, the rear hemisphere of the sphere was marked, reflecting a reversed viewing vector from the system's origin. Conversely, when participants engaged from outside the building, the front half of the sphere was marked. This map reveals distinct shifts in user focus based on the availability of feedback.

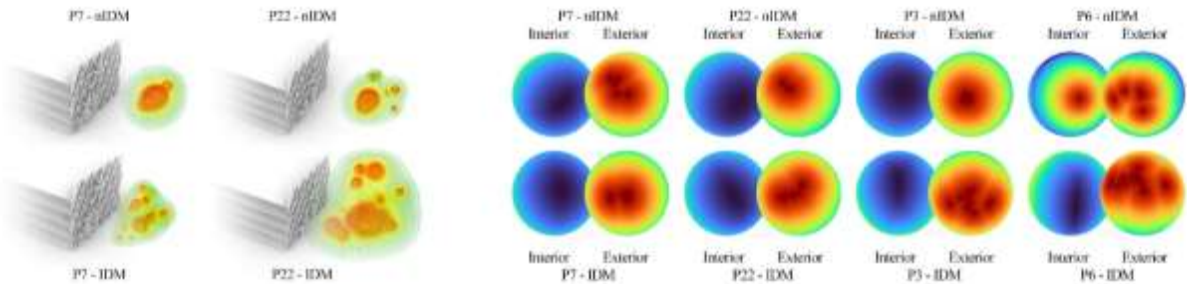

Figure 5: Left, distribution of sample players' movement while configuring the design; Right, distribution of players' focal points and stance relative to the design subject.

Additionally, participants who received feedback engaged more extensively with the design environment. They made significantly more design alterations and explored a greater number of unique configurations compared to those in the no-feedback condition, demonstrating heightened engagement and interaction. While feedback did not significantly affect users' viewing angles, it did result in broader positional movement, suggesting that users physically navigated the design space more actively when receiving performance input (Figure 5). This indicates that the framework effectively promoted spatial exploration without altering interaction directionality.

## 4. Conclusions

This study demonstrates that real-time, encapsulated feedback significantly enhances the design behavior of non-expert users across multiple dimensions. Participants interacting under the Informed Decision-Making (IDM) condition not only made faster decisions but also exhibited signs of adaptive learning, as indicated by declining decision times over repeated interactions. More importantly, their final design outcomes showed measurable improvements across structural, environmental, and fabrication performance metrics, even though the underlying data were abstracted into simplified feedback categories. These findings suggest that participants effectively internalized the information provided, using it to guide performance-oriented choices. Additionally, spatial engagement and interaction behaviors revealed that users receiving feedback explored more configurations and physically navigated the design space more extensively, without losing focus. Together, these results support the efficacy of the proposed framework in enabling early-stage designers to make more informed, performative decisions through minimal yet strategically structured guidance. This has significant implications for increasing accessibility to performance-driven design processes among broader, non-specialist user groups.

## Acknowledgements

We would like to acknowledge the financial support provided by DIALOG Architects and MITACS for funding this research. Our sincere thanks also go to all the participants who tested the framework and provided valuable feedback, contributing significantly to the success of this study.

## References

[1] T. Selker, "Fostering motivation and creativity for computer users," Int J Hum Comput Stud, vol. 63, no. 4–5, pp. 410–421, Oct. 2005, doi: 10.1016/j.ijhcs.2005.04.005.

[2] A. Khalilbeigi Khameneh and A. Nahmad-Vazquez, "A Gamified Framework for Non-Expert Decision-Making," in Designing Change: Proceedings Volume 1 For The 2024 Association For Computer Aided Design In Architecture Conference, A. Nahmad-Vazquez, J. Johnson, Taron. Joshua, J. Rhee, and D. Hapton, Eds., Banff: ACADIA, Nov. 2024, pp. 545–557.

[3] A. Mostafavi, "Architecture, biometrics, and virtual environments triangulation: a research review," Archit Sci Rev, pp. 1–18, Dec. 2021, doi: 10.1080/00038628.2021.2008300.

[4] Y. Friedman, "The Flatwriter: Choice by Computer," Prog Arch, vol. 98?101, Mar. 1971, [Online]. Available: https://usmodernist.org/PA/PA-1971-03.pdf

[5] A. Vrcelj Božić, N. Hoić-Božić, M. Holenko Dlab, K. Stančin, and T. Jagušt, "Gamification Classification as a Framework for Analysing Gamification Elements in Digital Educational Tools," Journal of Communications Software and Systems, vol. 21, no. 1, pp. 34–42, 2025, doi: 10.24138/jcomss-2024-0111.

[6] Mathias. Fuchs, Sonia. Fizek, and Paolo. Ruffino, Rethinking Gamification. Leuphana Universität Lüneburg. Innovations-Inkubator - meson Press by Hybrid Publishing Lab, 2014.

[7] K. Seaborn and D. I. Fels, "Gamification in theory and action: A survey," Int J Hum Comput Stud, vol. 74, pp. 14–31, Feb. 2015, doi: 10.1016/j.ijhcs.2014.09.006.

[8] D. B. Clark, E. E. Tanner-Smith, and S. S. Killingsworth, "Digital Games, Design, and Learning," Rev Educ Res, vol. 86, no. 1, pp. 79–122, Mar. 2016, doi: 10.3102/0034654315582065.

[9] R. S. Alsawaier, "The effect of gamification on motivation and engagement," The International Journal of Information and Learning Technology, vol. 35, no. 1, pp. 56–79, Jan. 2018, doi: 10.1108/IJILT-02-2017-0009.

[10] D. A. Norman and S. W. Draper, User centered system design: new perspectives on human-computer interaction. CRC Press, an imprint of Taylor and Francis, 1986.

[11] E. Di Matteo, P. Roma, S. Zafonte, U. Panniello, and L. Abbate, "Development of a Decision Support System Framework for Cultural Heritage Management," Sustainability, vol. 13, no. 13, p. 7070, Jun. 2021, doi: 10.3390/su13137070.

[12] E. Azar et al., "Simulation-Aided Occupant-Centric Building Design: A Critical Review of Tools, Methods, and Applications," Energy Build, vol. 110292 224 :1?17, Oct. 2020, doi: 10.1016/j.enbuild.2020.110292.

[13] P. Wacnik, S. Daly, and A. Verma, "Participatory design: A systematic review and insights for future practice," Sep. 2024. Accessed: May 20, 2025. [Online]. Available: https://arxiv.org/pdf/2409.17952

[14] C. Schultz, M. Bhatt, and A. Borrmann, "Bridging Qualitative Spatial Constraints and Feature-Based Parametric Modelling: Expressing Visibility and Movement Constraints," Advanced Engineering Informatics, vol. 31 :2?17, Jan. 2017, doi: 10.1016/J.AEI.2015.10.004.

[15] S. D. Kleban, W. A. Stubblefield, K. W. Mitchiner, J. L. Mitchiner, and M. Arms, "Collaborative Evaluation of Early Design Decisions and Product Manufacturability," in Proceedings of the Hawaii International Conference on System Sciences, 270, 2001. doi: 10.1109/HICSS.2001.927226.

[16] D. Norman, The Design of Everyday Things. The Design of Everyday Things, 2016. doi: 10.1 5358/9783800648108.

[17] B. Myers, S. E. Hudson, and R. Pausch, "Past, Present, and Future of User Interface Software Tools," ACM Transactions on Computer-Human Interaction, vol. 7, no. 1, 2000, doi: 10.1145/344949.344959.

[18] S. D'oca, H. B. Gunay, S. Gilani, and W. O?brien, "Critical Review and Illustrative Examples of Office Occupant Modelling Formalisms," Building Serv. Eng. Res. Technol., vol. 40, no. 6, pp. 732–757, 2019, doi: 10.1177/0143624419827468.

[19] J. Mahecha Zambrano, U. F. Oberegger, and G. Salvalai, "Towards Integrating Occupant Behaviour Modelling in Simulation-Aided Building Design: Reasons, Challenges and Solutions," Energy Build, vol. 253 :111498, Dec. 2021, doi: 10.1016/J.ENBUILD.2021.111498.

[20] C. Preisinger, M. Heimrath, M. Arch, and B. G. Schneider, "Karamba," A Toolkit for Parametric Structural Design.? Structural Engineering International, vol. 24, no. 2, pp. 217–221, 2014, doi: 10.2749/101686614X13830790993483.

[21] A. Elesawy, S. Caranovic, J. Zarb, P. Jayathissa, and A. Schlueter, "HIVE Parametric Tool - A Simplified Energy Simulation Tool for Educating Architecture Students," in Computing for a Better Tomorrow - Proceedings of the 36th ECAADe Conference - Volume 1, Lodz University of Technology, Lodz, Poland, 19-21 September 2018, 1:657?66, A. Kepczynska-Walczak and S. Bialkowski, Eds., 2018, pp. 657–666. doi: 10.52842/CONF.ECAADE.2018.1.657.

[22] R. A. Grier, A. Bangor, P. Kortum, and S. C. Peres, "The System Usability Scale," in Proceedings of the Human Factors and Ergonomics Society Annual Meeting, September, 187?91, 2013. doi: 10.1177/1541931213571042.